\documentclass[english,showpacs,preprintnumbers,amsmath,amssymb,12pt]{revtex4}
\usepackage[]{fontenc}
\usepackage[latin1]{inputenc}

\makeatletter

\usepackage{graphicx}
\usepackage{dcolumn}
\usepackage{bm}

\usepackage{babel}
\makeatother
\begin{document}

\title{Spin flip scattering at Al surfaces.}

\author{N. Poli}

\email{poli@kth.se}

\author{M. Urech, V. Korenivski, D. B. Haviland}

\affiliation{Nanostructure Physics, Royal Institute of Technology, Albanova University
Centre, SE-10691 Stockholm, Sweden}

\begin{abstract}
Non-local measurements are performed on a multi terminal device to
$in-situ$ determine the spin diffusion length and in combination
with resistivity measurements also the spin relaxation time in Al
films. By varying the thickness of Al we determine the contribution to spin relaxation from surface scattering. From the temperature dependence
of the spin diffusion length it is established that the spin relaxation
is impurity dominated at low temperature. A comparison of the spin
and momentum relaxation lengths for different thicknesses reveals
that the spin flip scattering at the surfaces is weak compared to
that within the bulk of the Al films. 
\end{abstract}

\maketitle
The first experiment on spin injection and detection in metals dates
back to 1985 when Johnson and Silsbee \cite{Lubzens,Johnson:85} demonstrated
spin accumulation in a single-crystal Al for temperatures below 77
K, with a spin diffusion length, $\lambda_{sf}\sim100\ \mu m$. Recent
experiments on thin films \cite{Bass,Jedema:01,Jedema:02,Jedema:03,Urech}
found orders of magnitude shorter $\lambda_{sf}\sim0.1-1\ \mu$m.
However, the measured spin accumulation is greatly enhanced due to much reduced effective device volumes. Understanding the origins of spin relaxation in such devices is therefore important for spintronics applications. The main contribution to spin relaxation in a metal is the spin orbit coupling induced by the lattice ions in the metal \cite{Elliot,Yafet}. In combination with momentum scattering, spin flip events can occur with a certain (typically small) probability, which depends on the specific band structure, nature of the impurities present, phonons, etc. As the dimensions of electronic devices decrease, the influence of surface scattering becomes important. We analyze
this contribution by studying the temperature and thickness dependence
of spin relaxation parameters in thin Al films. We find that the spin
relaxation is impurity dominated at low temperatures ($LT$) and,
surprisingly, that the spin flip scattering at the surfaces is negligible
compared to that within the bulk. 

Multi terminal lateral devices allow several simultaneous measurements
to be conducted. For example, the spin imbalance in a non-magnetic
metal (NM) produced by current injection from a ferromagnet (FM) can
be detected at various distances from the injection point using a
set of magnetic electrodes, as illustrated in Figure 1. Thus, detecting voltages
non-locally (outside the current path) at two or more locations of the Al strip allows an in-situ
determination of $\lambda_{sf}$ \cite{Urech}. The non-equilibrium
spin accumulation produced at the injection point diffuses along the
Al, governed by the diffusion equation, \begin{equation}
\nabla^{2}\delta\mu=\delta\mu/\lambda_{sf}^{2},\label{DiffusionEquation}\end{equation}
 where $\delta\mu=(\mu_{\uparrow}-\mu_{\downarrow})$ is the spin
splitting in the chemical potential, $\lambda_{sf}=\sqrt{D\tau_{sf}}$,
and $D$ is the diffusion constant given by the Einstein relation,
$\sigma=e^{2}ND$. The solution of (\ref{DiffusionEquation}) with appropiate boundary conditions gives the \char`\"{}spin signal\char`\"{}
at a distance $x$ from the injection point \begin{equation}
R_{x}=\frac{\Delta V}{I_{inj.}}=P^{2}\frac{\rho\lambda_{sf}}{A}e^{-x/\lambda_{sf}},\label{Rx}\end{equation}
 where $P$ is the polarization of the injector, $\rho$ and
$A$ are the resistivity and the cross sectional area of the NM, respectively. 

The resistivity of a metal is expected to increase rapidly when the
thickness of the metal becomes comparable to or smaller than the mean
free path. This increase is due to scattering at surfaces \cite{Fuchs,Sondheimer,Soffer}
and grain boundaries \cite{Mayadas}. These two contributions were
considered by Sambles' \cite{Sambles} who assigned an angle dependent specularity $p_{i}(\cos\theta)$ to each surface and assumed columnar grain growth, with the average grain
diameter $D$ and grain boundary specularity $R$. The result of this
model for the ratio of the bulk resistivity $\rho_{0}$ to the total
resistivity $\rho_{\textrm{tot}}$ can be expressed as a function of the film
thickness normalised to the bulk mean free path ($k=d/\lambda_{0}$) \begin{equation}
\begin{split} & \frac{\rho_{0}}{\rho_{tot}}=G(\alpha)-\frac{4}{\pi}\int_{0}^{\frac{\pi}{2}}d\phi\int_{0}^{1}du\cos^{2}\phi~3(u-u^{3})\times\\
\\ & \frac{\{1-exp(-kH/u)\}\{1-\bar{p}+(\bar{p}-p_{1}p_{2})exp(-kH/u)\}}{2kH^{2}\{1-p_{1}p_{2}\, exp(-2kH/u)\}},\end{split}
\label{rho_tot}\end{equation}
 where \begin{eqnarray*}
G(\alpha) & = & 1-\frac{3}{2}\alpha+3\alpha^{2}-3\alpha^{3}\ln\left(1+\frac{1}{\alpha}\right),\\
H & = & 1+\frac{\alpha}{\sqrt{(1-u^{2})}\cos\phi},\end{eqnarray*}

\noindent $\alpha=R/(1-R)\cdot l_{0}/D$ and $\bar{p}=\frac{1}{2}(p_{1}+p_{2})$.
Using this model, we analyze the relative contributions of the surface
and grain boundary scattering to the resistivity of our Al films.

The samples were fabricated using electron beam lithography and the standard two angle deposition technique in an e-gun evaporation system. First, 40 $\mu$m long and 100 nm wide Al strips with different thicknesses were deposited at normal incidence, followed by oxidation in $O_{2}$ at a pressure of 80 mTorr for 8 min. Oxidation was performed at $RT$, providing an $Al_2O_{3}$ tunnel barrier with a typical specific resistance of $\sim 0.15$ k$\Omega\mu m^2$. Next, Co electrodes of 50-80 nm in thickness were deposited from an
angle of $40^{\circ}$ to overlap the Al strip. The Co electrodes
are designed to have different widths, between 60 and 70 nm, which results
in different coercive fields, allowing us to switch their magnetization
independently. Figure 1 shows a SEM micrograph of the device consisting
of 3 vertical Co electrodes, each separated by a distance of 300 nm,
overlapping an oxidized Al strip. 

To isolate the spin signal, non-local measurements \cite{Johnson:85}
were performed according to the electrical arrangement shown in Figure
1. This allowed us to \textit{in-situ} determine $\lambda_{sf}$ and,
combined with four point resistivity measurements, also $\tau_{sf}$.
A bias current of $1\ \mu$A was injected into the Al strip and the
voltages at the detectors 1 and 2 outside the current path were measured
simultaneously using a standard lock-in technique at 7 Hz. High input
impedance ($\sim10^{15}\ \Omega$) voltage pre-amplifiers with low
input bias current (1-10 fA) were used to minimize the noise and current
leakage. An external magnetic field was applied in the plane along the length of
the FM electrodes in order to switch the magnetizations to the desired states.
The measurements were performed for a set of samples having different
Al thickness.

Figure 2 shows the spin signals at the two detectors as a function
of the external magnetic field for a sample with 15 nm thick Al measured
at 4 K. To begin with, all the electrodes were saturated in the negative
direction ($\downarrow\downarrow\downarrow$) and then switched separately
by ramping the field in the positive direction. First, detector 1
switches ($\downarrow\uparrow\downarrow$) at 1240 Oe, then detector
2 ($\uparrow\uparrow\downarrow$) at 1675 Oe. Finally the injector switches
($\uparrow\uparrow\uparrow$) at 1840 Oe, which one can see as a simultaneous transition in both curves of Figure 2. The measured spin signals at two distances, together with equation \ref{Rx} gives $P=12\%$ and $\lambda_{sf}=660$ nm. Note that both properties
were obtained \textit{in-situ} in one field sweep, which eliminates
uncertainties due to irreproducibilities in fabrication. We note, that this P value is the effective spin polarization of the injecting interface and is not equal to the bulk polarization of Co (see also \cite{Tinkham}).

The temperature dependence of $\lambda_{sf}$ for the sample with 15 nm thick Al strip is shown in Figure 3. As $T$ is lowered $\lambda_{sf}$ increases
from $\approx350$ nm at $RT$ until it levels off at $\approx660$
nm at $LT$, which demonstrates that the spin relaxation is impurity
dominated at $LT$. The data reveals a contribution from phonon mediated
scattering, which significantly reduces $\lambda_{sf}$ for $T>50$ K. The analysis of the temperature dependence of the resistivity (not shown) reveals that both the diffusion constant and spin relaxation time increase with lowering temperature. Our results on the temperature dependence can be reconciled with
the early single crystal data \cite{Lubzens,Johnson:85} using the
theory of \cite{FabianSarma:99}. The sole difference appears to be
the greater amount of impurities in thin films. 

Figure 4 shows the resistivity of the Al films measured at 4 K as
a function of thickness together with a theoretical fit according
to Eq. \ref{rho_tot}. The line represents the theoretical prediction of $\rho_{\textrm{tot}}$ for the grain size $D\approx6$ nm, which was determined by AFM measurements of the topography of the films. The fact that the grain size essentially does not change with thickness results in approximately constant background to the total resistivity from grain boundary scattering. Thus, the dominating contribution to the resistivity of the thinnest films is
the diffusive scattering at the surfaces. 

To investigate how the additional surface momentum scattering affects
spin relaxation, the thickness dependence of $\lambda_{sf}$ and $\lambda_{p}$
at 4 K was determined, where $\lambda_p$ is the thickness dependent mean free path determined by the measured resistivity. In the bulk limit, where the thickness is greater than both $\lambda_{sf}$ and $\lambda_{p}$, the ratio of the two should be constant and is estimated to be $\approx15$. The bulk $\lambda_{0}$ is obtained by extrapolation in Figure 4 to large thicknesses, and
the bulk $\lambda_{sf}$ is estimated from the analysis of the spin
relaxation time (to be published elsewhere). In thin films, surface
scattering determines $\lambda_{p}$. If the surface scattering contribution to spin relaxation is equally strong as that of the bulk impurity scattering, then the ratio of the two characteristic lengths should be independent of thickness. However, the measured data for $\lambda_{sf}/\lambda_{p}$, plotted in Figure 5,  shows a clear increase for small thickness. This means that the spin flip scattering at the surfaces is weak compared to the spin relaxation within the bulk of the films. Seen from a persective of "spin hot spots"

Spin relaxation in Al sensitively depends on the details of the band structure (the so-called {}``spin hot spots'', \cite{FabianSarma:99}). It is then natural to expect that the bulk electronic structure is significantly
perturbed at the surfaces in such a way that the spin-flip scattering
cross-sections are reduced. This provides a qualitative explanation
for the weak spin relaxation at the surface we observe.

In conclusion, we have measured the spin relaxation length in Al films
as a function of temperature and thickness. We observe that the spin
relaxation is dominated by impurity scattering at $LT$. Interestingly,
the contribution from surface scattering to spin relaxation is found
to be weak compared to spin relaxation within the bulk of the film. 

N. Poli and M. Urech gratefully acknowledge support from the Swedish
SSF. We thank John Slonczewski
for fruitful discussions.

\newpage
\section*{Figure captions}

\begin{enumerate}
\item Scanning electron micrograph of a multi terminal device. The vertical
Co electrodes are in contact with the Al strip through tunnel barriers
($Al_2O_{3}$).
\item Spin voltages versus applied magnetic field for a 15 nm Al sample
measured at 4K: at detector 1 (upper panel); detector 2 (lower panel).
The arrows indicate the magnetic states of the electrodes.
\item Spin diffusion length, $\lambda_{sf}$, as a function of temperature
for a 15 nm Al sample. At high temperatures, the dominant scattering mechanism is with phonons, whereas it is with impurities at $LT$ ($T<50$ K).
\item Resistivity of Al films as a function of thickness. The solid line
is a theoretical fit for $\rho_{\textrm{tot}}$ with $D\approx6$ nm, $R\approx 0.1$, $\rho_0\approx 0.2~\mu\Omega$ cm, $\lambda_0\approx200$ nm and maximal surface diffusivities.
\item The ratio of the spin diffusion length and the momentum mean free
path as a function of the Al film thickness at 4 K.
\end{enumerate}

\newpage
\begin{figure}[h!]  
        \centering
                \includegraphics[width=0.6\textwidth]{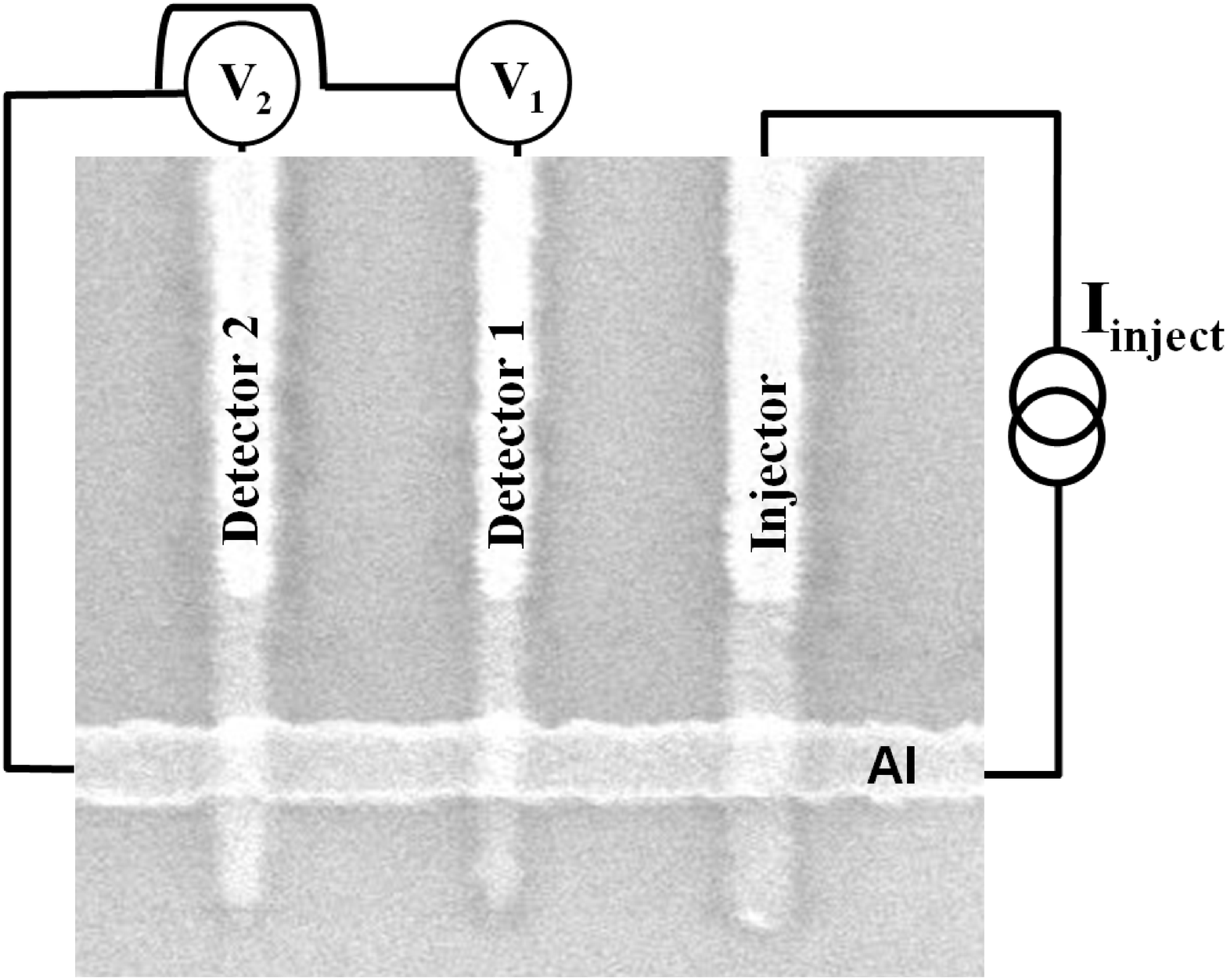}
                \caption{Poli et Al}\label{Switch1&2}
\end{figure}
\begin{figure}[h!]  
        \centering
                \includegraphics[width=0.6\textwidth]{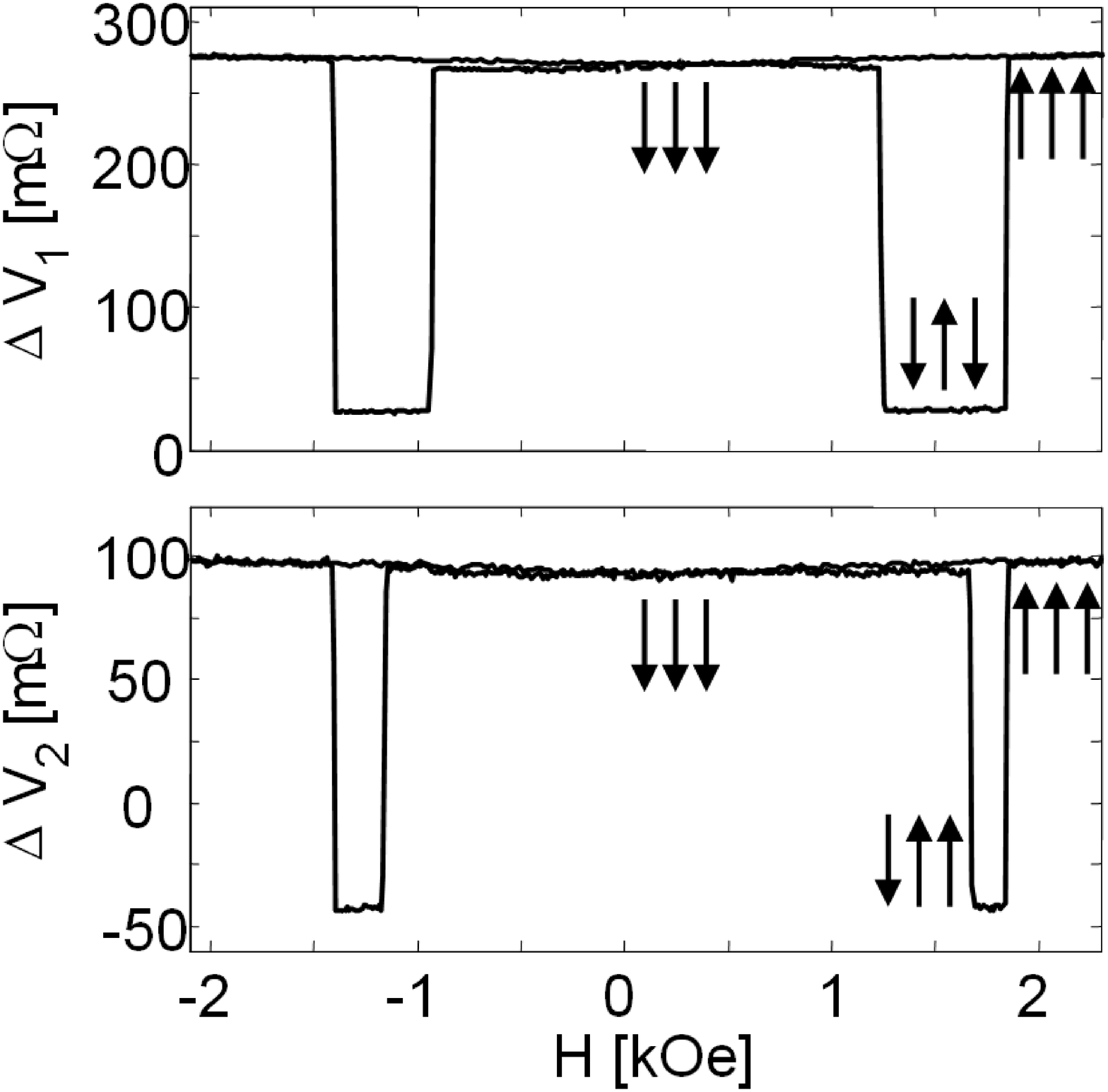}
                \caption{Poli et Al}\label{Switch1&2}
\end{figure}

\begin{figure}[h!]  
        \centering
                \includegraphics[width=0.8\textwidth]{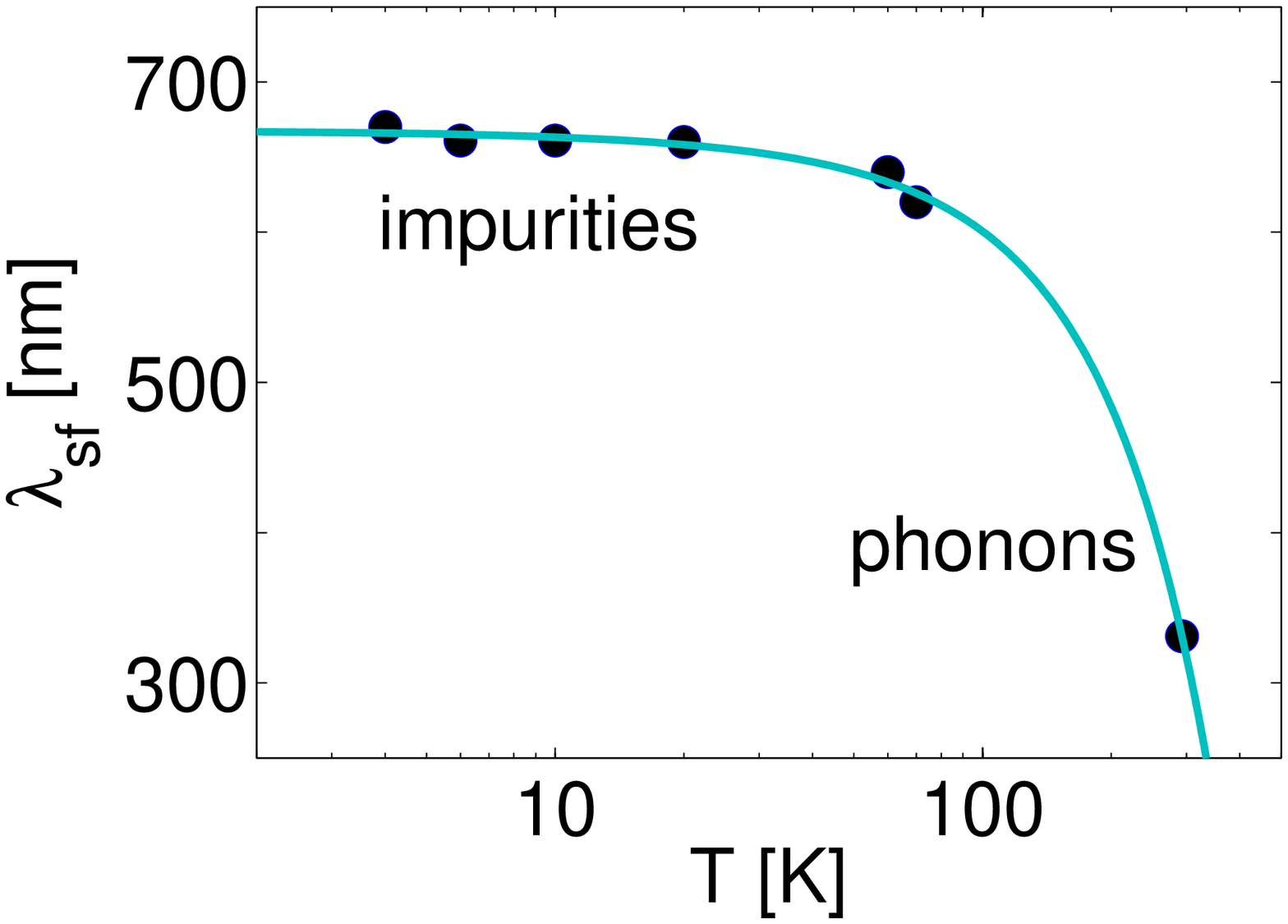}
                \caption{Poli et Al}\label{Switch1&2}
\end{figure}
\begin{figure}[h!]  
        \centering
                \includegraphics[width=0.75\textwidth]{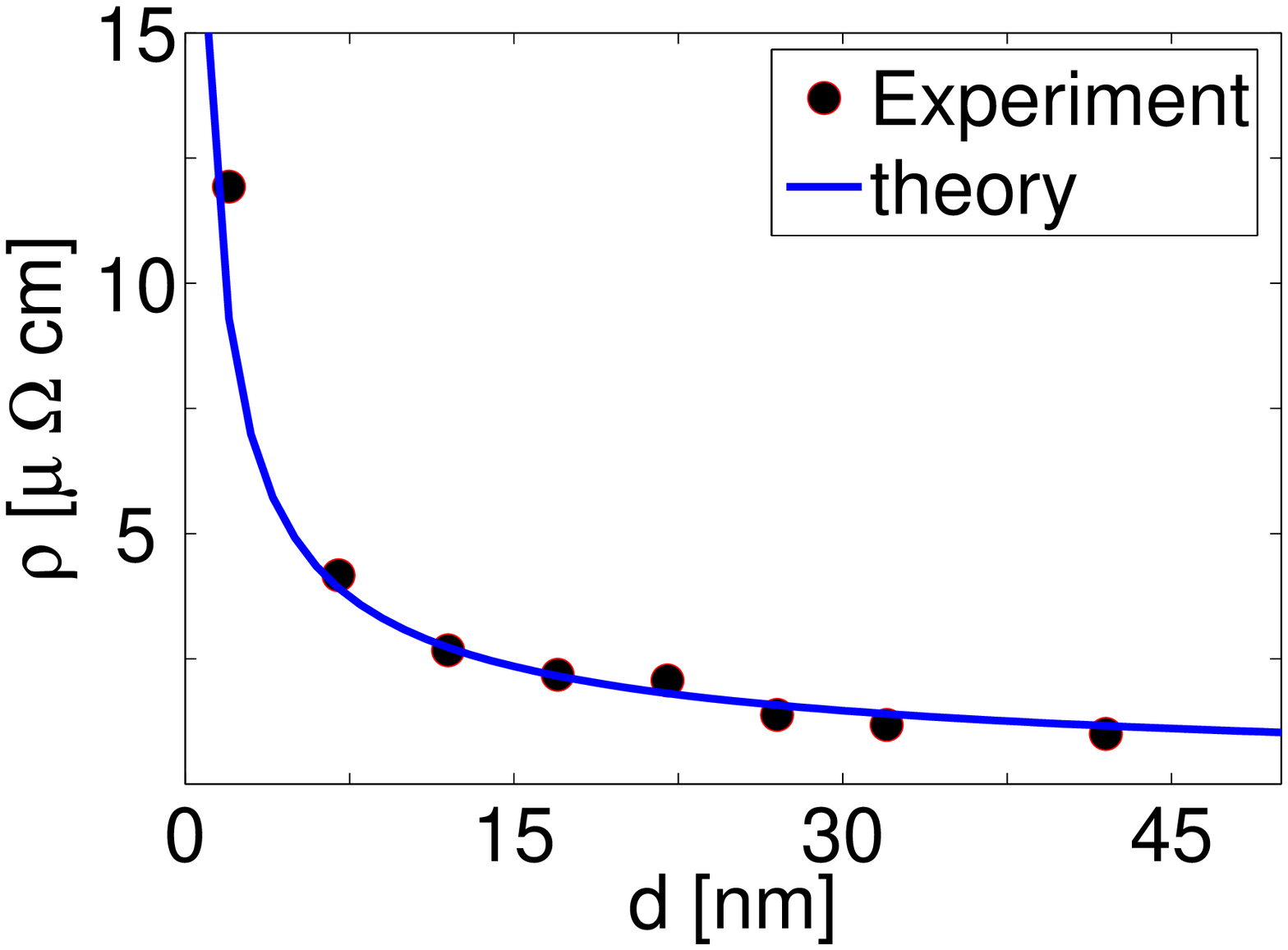}
                \caption{Poli et Al}\label{Switch1&2}
\end{figure}
\begin{figure}[h!]  
        \centering
                \includegraphics[width=0.8\textwidth]{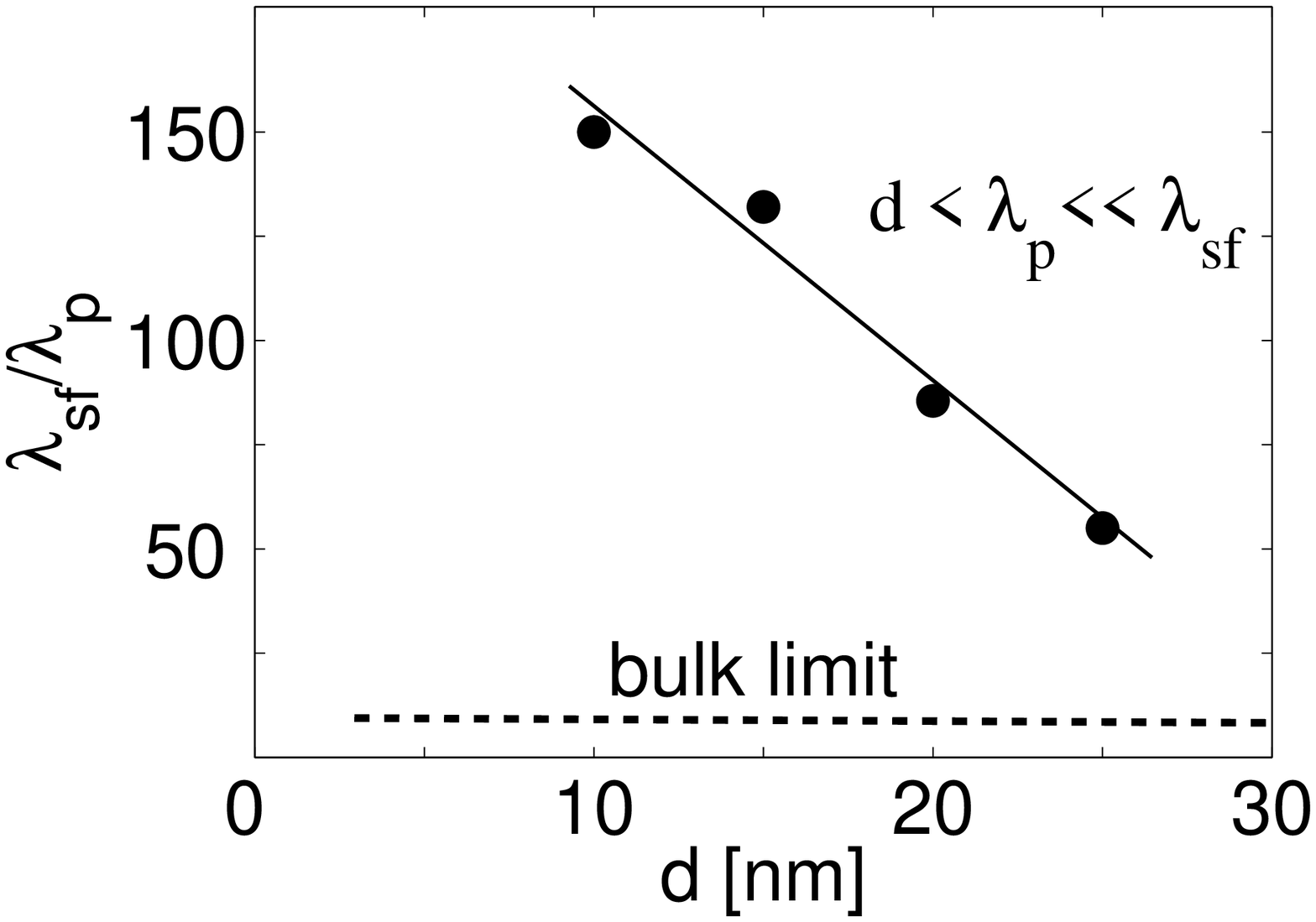}
                \caption{Poli et Al}\label{Switch1&2}
\end{figure}


\newpage

\begin{thebibliography}{10}
\bibitem{Lubzens}D. Lubzens and S. Schultz, Phys. Rev. Lett., \textbf{36}, 1104 (1976) 
\bibitem{Johnson:85}M. Johnson and R. H. Silsbee, Phys. Rev. Lett., \textbf{55}, 1790
(1985) 
\bibitem{Bass}J. Bass and W. P. Pratt, J. Magn. Magn. Mater. \textbf{200}, 274 (1999). 
\bibitem{Jedema:01}F. J. Jedema \textit{et al}, Nature, \textbf{410}, 345 (2001) 
\bibitem{Jedema:02}F. J. Jedema \textit{et al}, Nature, \textbf{416}, 713 (2002) 
\bibitem{Jedema:03}F. J. Jedema \textit{et al}, Phys. Rev B, \textbf{67}, 085319 (2003) 
\bibitem{Elliot}R. J. Elliot, Phys. Rev. \textbf{96}, 266 (1954). 
\bibitem{Yafet}Y. Yafet, in \textit{Solid state physics}, \textbf{14}, (1963) 
\bibitem{Fuchs}K. Fuchs, Proc. Cambridge Philos. Soc. \textbf{34}, p. 100 (1938) 
\bibitem{Sondheimer}E. H. Sondheimer, Adv. Phys. \textbf{1}, p. 1 (1952) 
\bibitem{Soffer}S. B. Soffer, J. Appl. Phys. \textbf{38}, 1710 (1967) 
\bibitem{Mayadas}A. F. Maydas and M. Shatzkes, Phys. Rev. B \textbf{1}, 1382 (1970) 
\bibitem{Sambles}J. R. Sambles, Thin Solid Films \textbf{106}, 321 (1983) 
\bibitem{FabianSarma:99}J. Fabian and S. Das Sarma, Phys. Rev. Lett., \textbf{83}, 1211 (1999) 
\bibitem{Urech}M. Urech, J. Johansson, V. Korenivski and D. B. Haviland, J. Magn.
Magn. Mater. \textbf{272-276}, e1417 (2004) 
\bibitem{Tellier}C. R. Tellier and A. J. Tosser, Thin Solid Films \textbf{37}, 207
(1976) 
\bibitem{Tinkham}S. O. Valenzuela and M. Tinkham, App. Phys. Lett. \textbf{85}, 5914
(2004) 
\end{thebibliography}
\end{document}